\documentclass[letter,11pt]{article}
\pdfoutput=1 % if your are submitting a pdflatex (i.e. if you have
\usepackage{jcappub} % for details on the use of the package, please
\usepackage[T1]{fontenc} % if needed
\usepackage{cleveref}
\usepackage{graphicx}
\usepackage{caption}
\usepackage{subcaption}
\usepackage{ulem} % for \sout
\usepackage{bm}

\def\bdm{\begin{displaymath}}
\def\edm{\end{displaymath}}

\def\barray{\begin{array}}
\def\earray{\end{array}}
\def\be{\begin{equation}}
\def\ee{\end{equation}}
\def\ben{\begin{equation} \nonumber}
\def\een{\end{equation}}
\def\ban{\begin{eqnarray*}}
\def\ean{\end{eqnarray*}}
\def\ba{\begin{eqnarray}}
\def\ea{\end{eqnarray}}
\def\eal{\end{align}}
\def\bal{\begin{align}}

\def\({\left(}
\def\){\right)}
\def\[{\left[}
\def\]{\right]}

\def\ep{\epsilon}

\def\b{\beta}

\def\by{{\bf{y}}}
\def\ba{{a\left(\by\right)}}

\def\bk{{\bf k}}

\def\bx{{\bf x}}

\begin{document}

\title{The onset of the strong backreaction regime in axion inflation, an analytical study}

\author{Aditya Kulkarni}
\affiliation{Amherst Center for Fundamental Interactions, Department of Physics, University of Massachusetts, Amherst, MA 01003, U.S.A.}
\author{and Lorenzo Sorbo}

% The "\note" macro will give a warning: "Ignoring empty anchor..."
% you can safely ignore it.

% e-mail addresses: one for each author, in the same order as the authors

\emailAdd{abkulkarni@umass.edu}
\emailAdd{sorbo@umass.edu}

\abstract{Models where an axion-like inflaton is coupled to a $U(1)$ gauge field are theoretically well motivated and can display a rich phenomenology. The regime in which the quanta of the gauge field strongly backreact on the rolling inflaton has been studied for well more than a decade, and yet we cannot say that it is fully understood.  In this paper we present an analytical study of the onset of the strong backreaction regime. Our formalism relies on a Laplace transform to convert a complicated integro-differential equation into the search of the poles of an analytic function. We confirm the existence, previously observed using a completely different formalism, of a stable region of strong backreaction. Our work provides, for the first time, analytical formulae that allow to study the evolution of the system as it transitions from the weak to the strong backreaction regime.}

\maketitle


%%%%%%%%%%%%%%%%%%%%%%%
\section{Introduction}%
%%%%%%%%%%%%%%%%%%%%%%%

Any predictive model of inflation must explain why the inflaton potential remains sufficiently flat once radiative corrections are taken into account. This requirement is quantified by the smallness of the  slow-roll parameters $\epsilon\equiv M_{\rm P}^2\,V'{}^2/(2\,V^2)$ and $\eta \equiv M_{\rm P}^2\, V''/V$. The fact that these quantities, for a renormalized scalar field, are generally not small is known as the ``$\eta$ problem''. The $\eta$ problem can be evaded if the inflaton interacts only with itself and gravity~\cite{Linde:1987yb,Kaloper:2008fb}, but decoupling the inflaton from all other sectors is not a satisfactory solution, since the inflaton must ultimately transfer its energy to Standard Model degrees of freedom in order to reheat the Universe. A more robust strategy is to protect the flatness of the potential by requiring that the inflaton enjoys a softly broken shift symmetry~\cite{Freese:1990rb}. In this case the inflaton is typically a pseudoscalar degree of freedom, which we will refer to as an {\it axion}. Although the  cosine potential originally proposed in~\cite{Freese:1990rb} is now disfavored by CMB data~\cite{Planck2018,BICEPKeck2021}, a variety of non-minimal realizations such as, e.g.,~\cite{Kim:2004rp,Peloso:2015dsa,DAmico:2017cda,DAmico:2018mnx}, remain compatible with observations. 

Crucially, the same shift symmetry that protects the potential also allows a dimension-five, parity-odd 
coupling of the axion to gauge fields\footnote{The shift symmetry also allows efficient interactions of the axion with fermions, leading to fermion production during inflation as well as in more general cosmological backgrounds~\cite{Adshead:2015jza,Adshead:2015kza,Adshead:2018oaa,Domcke:2018eki,Adshead:2019aac,Roberts:2021plm,Kulkarni:2024kzc}.} of the Chern--Simons form, $ \mathcal{L} \supset -\,\frac{\alpha}{4 f}\,\Phi\,F_{\mu\nu}\tilde{F}^{\mu\nu},$ where $\alpha/f$ is a coupling constant with dimensions of length and $F_{\mu\nu}$ is the field strength of an Abelian gauge field.\footnote{Analogous structures have been studied for non-Abelian sectors (see e.g.~\cite{Adshead:2012kp,Domcke:2018rvv}) and for spectator axions (see e.g.~\cite{Barnaby:2012xt,Namba:2015gja}).} It has been shown~\cite{Anber:2006xt} that this interaction induces a tachyonic amplification of the gauge field modes which, in its turn, leads to a very rich phenomenology. The gauge fields produced during inflation may survive as the seeds of the large-scale magnetic fields observed today~\cite{Turner:1987bw,Garretson:1992vt,Anber:2006xt,Adshead:2016iae,Sobol:2019xls}, or, if the gauge group is a hidden $U(1)$, constitute a viable dark-photon dark-matter candidate~\cite{Bastero-Gil:2021wsf,Domcke:2021yuz}. Phenomenological predictions also include nongaussianities~\cite{Barnaby:2010vf}, deviations from scale invariance~\cite{Namba:2015gja}, formation of a population of primordial black holes~\cite{Linde:2012bt}, generation of primordial chiral gravitational waves at CMB~\cite{Sorbo:2011rz} or interferometer~\cite{Cook:2011hg,Corba:2024tfz,Corba:2025reo} frequencies and baryogenesis~\cite{Anber:2015yca}. See~\cite{Pajer:2013fsa} for a review.

For small values of $\alpha/f$, the gauge field can be treated as a spectator and the production of its quanta can be followed mode by mode in Fourier space. Once $\alpha/f$ becomes large enough, however, the source term $\propto \langle F_{\mu\nu}\tilde{F}^{\mu\nu}\rangle$ appearing in the axion equation of motion becomes comparable to the potential gradient $V'$, and the system enters the strong-backreaction regime, in which all modes are nonlinearly coupled. It was originally expected~\cite{Anber:2009ua} that the friction induced by strong backreaction could balance the potential gradient and sustain a {\it smooth} period of slow roll inflation even on steep potentials in what has been referred to as the ``the Anber--Sorbo (AS) solution''. However, it was later realized~\cite{Cheng:2015oqa,Notari:2016npn,DallAgata:2019yrr,Domcke:2020zez,Caravano:2022epk,Gorbar:2021rlt}  that the response of the gauge field to changes in the inflaton velocity $\dot{\Phi}$ is typically delayed~\cite{Notari:2016npn,Domcke:2020zez} by ${\cal O}(1)$ efoldings, and this lag drives large oscillations in time of both $\dot{\Phi}$ and the gauge-field energy density.\footnote{Conversely, ref.~\cite{Creminelli:2023aly} argued that, in systems in which the response of the produced matter field to changes in $\dot\Phi$ occurs in much less than an efolding, those oscillations should not emerge, and presented a model where this situation is realized. More recently, ref.~\cite{Baker:2026xsd} has shown that a variation of the model discussed in the present paper, in which the gauge field is sufficiently massive, provides another example of a system where the mechanism of~\cite{Creminelli:2023aly} is at work and oscillations in $\dot\Phi$ do not emerge.} Such large oscillations are expected to leave an imprint in the primordial spectrum of scalar and~\cite{Garcia-Bellido:2023ser} tensor metric perturbations. Subsequent stability analyses concluded that the emergence of such oscillations is due to a rather generic instability of the AS solution~\cite{Peloso:2022ovc,vonEckardstein:2023gwk}.  This result was confirmed by lattice simulations which show that the large oscillations in $\dot{\Phi}$ are followed by a rapid growth of axion inhomogeneities~\cite{Figueroa:2023oxc,Figueroa:2024rkr} (see also~\cite{Sharma:2024nfu,Lizarraga:2025aiw,Iarygina:2025ncl} for additional lattice studies). More recently, a comprehensive scan of parameter space using the gradient expansion formalism~\cite{Sobol:2020lec,Gorbar:2021rlt} has identified a region in which the background stationary solution remains stable despite strong backreaction~\cite{Sobol:2026abc}. Even more recently, it has been shown~\cite{Domcke:2026fjo} that for sufficiently large values of $\alpha/f$, the system can reach a ``mild backreaction regime'' where it diverges significantly from the no-backreaction solution while evolving along the AS solution  for several efoldings before the appearance of large oscillations in $\dot\Phi$.  Remarkably,~\cite{Domcke:2026fjo} shows that along this solution spatial gradients of the inflaton can be kept under control.

Since the origin of the oscillations in $\dot\Phi$ during the strong backreaction epoch originates from the delayed backreaction effect of the produced matter, in this paper (as was already done in~\cite{Baker:2026xsd}) we will refer to what has been called AS solution as the ``instant backreaction'' (IBR) solution, a term that we find more descriptive of this regime.

In the present work we develop an analytical description of the departure from the weak backreaction regime and the onset of the backreaction--dominated dynamics. Analogously to what was made in~\cite{Peloso:2022ovc}, we take the IBR solution as our reference background and systematically expand the coupled axion--gauge system around it, retaining the leading corrections associated with the exact evolution of the inflaton, and, in addition to what was made in~\cite{Peloso:2022ovc}, of the scale factor and the Hubble parameter $H$. This allows us to follow how small departures from the IBR regime modify both the homogeneous axion evolution and the amplified gauge-field modes, and ultimately to determine when these departures grow into an instability. Keeping the leading corrections associated with the slow roll evolution of inflaton, scale factor and Hubble parameter produces source terms that displace the system from the stationary configuration. This should be contrasted with the analysis of~\cite{Peloso:2022ovc}, where the source terms were ignored and the authors focused on determining the (in)stability of the system by computing its Lyapunov exponents.  We solve the linearized gauge field equation using its retarded Green's function and insert the resulting gauge field response back into the homogeneous axion equation, so that the dynamics is reduced to a linear, inhomogeneous integro-differential equation. The nonlocal structure of this equation takes a much simpler representation in Laplace space (see~\cite{Belrhali:2026ygh}  for a different, recent application of Laplace transforms in cosmology), where the convolution integrals factorize and the problem reduces to an algebraic equation. The subsequent evolution is  encoded directly in the singularity structure of the Laplace transform.

The Laplace representation makes the transition between qualitatively different dynamical regimes especially transparent. The linearized solution for $\Phi(t)$ about the IBR evolution contains a polynomial contribution, a decaying mode, and an infinite set of modes associated with complex conjugate zeros of the transcendental part of the Laplace transform. Analogously to what was found in~\cite{Peloso:2022ovc}, the late time behavior is controlled by the pole with the largest real part, which we denote by $\alpha_0$: perturbations decay for $\mathrm{Re}(\alpha_0)<0$ and grow for $\mathrm{Re}(\alpha_0)>0$. We use this criterion to construct the stability diagram in the $(\kappa,\,\xi)$ plane, where $\kappa$ (defined in eq.~(\ref{eq:defkappa}) below, see also eq.~(\ref{eq:alsokappa})) measures the strength of gauge field backreaction on the homogeneous axion evolution, while $\xi$ denotes, as is customary, the combination $\alpha\,|\dot\Phi|/(2\,f\,H)$. Our analysis corroborates the results that were obtained in~\cite{Sobol:2026abc} using the gradient expansion formalism, where it was noted that the onset of an instability does not coincide, in general, with the conventional boundary between weak and strong backreaction. In particular, we find a region with $\kappa<1$ in which the gauge field friction is still subdominant but the stationary solution is already unstable, as well as a region with $\kappa>1$ in which backreaction is strong while perturbations around the stationary solution remain stable. Furthermore, by expanding about the point at which $\mathrm{Re}(\alpha_0)$ first crosses zero, we obtain a first-order analytical expression for the evolution of $\xi$ at the onset of the instability. We check that, for a range of values of the coupling $\alpha/f$, this approximation reproduces quite accurately the leading departure from the stationary evolution observed in the direct numerical solution of the coupled homogeneous axion and gauge-field equations.

The remainder of this paper is organized as follows. In Section~\ref{sec:equations} we derive the exact equations governing the axion and the $U(1)$ gauge field on a Friedmann-Robertson-Walker (FRW) background. In Section~\ref{sec:solutions} we review the stationary IBR solution and construct the linearized system describing deviations from it, including the leading corrections to the inflaton, the Hubble rate, the scale factor, and the amplified gauge field modes; the latter are expressed in terms of an appropriate retarded propagator. In Section~\ref{sec:lap} we reduce the resulting system in the large $\xi$ and slow roll limits and solve the corresponding integro-differential equation by means of a Laplace transform. We analyze the poles of the transformed solution and evaluate the sign of the real part of the rightmost pole to determine the stability regions in the $(\kappa,\xi)$ plane. In Section~\ref{sec:num} we translate this solution into the evolution of the instability parameter $\xi$ and compare the semi-analytical prediction with a direct numerical evolution of the coupled homogeneous system. We summarize our results and discuss their implications in Section~\ref{sec:summary}. Technical details regarding the calculation of the Laplace transforms and a discussion on the effects of placing a hard UV cutoff on the momentum integral in our main equation are collected in two appendices.

%%%%%%%%%%%%%%%%%%%%%%%%%%%%%%%%%%%%%%%%%%%%%%%%%%%%%%%%%%%%%%%%%%%%%%%%
\section{Exact equations for axion inflation with a \texorpdfstring{$U(1)$}{U(1)} gauge field}%
\label{sec:equations}%%%%%%%%%%%%%%%%%%%%%%%%%%%%%%%%%%%%%%%%%%%%%%%%%%%
%%%%%%%%%%%%%%%%%%%%%%%%%%%%%%%%%%%%%%%%%%%%%%%%%%%%%%%%%%%%%%%%%%%%%%%%

Our system consists of a pseudoscalar inflaton $\Phi$ with potential $V(\Phi)$ and a $U(1)$ gauge field $A_\mu$ with field strength $F_{\mu\nu}$. The two fields interact via a Chern-Simons--like operator with coupling constant $\alpha/f$, which has dimensions of a length. The system is on a flat FRW background, so that in our ``mostly plus'' convention the metric in conformal time $\tau$ takes the form $g_{\mu\nu}=a^2(\tau)\,{\rm {diag}}(-1,\,1,\,1,\,1)$.

The equations of motion read
\begin{align}
&{\nabla}^{\mu}{\nabla}_{\mu}\Phi-\frac{dV(\Phi)}{d\Phi}=\frac{\alpha}{4f}\,F_{\mu\nu}\,\tilde F^{\mu\nu}\,,\nonumber\\
&{\nabla}_{\mu}F^{\mu\nu}=-\frac{\alpha}{f}\left({\pmb\nabla}_{\mu}\Phi\right)\tilde F^{\mu\nu}\,,\nonumber\\
&{\nabla}_{\mu}\tilde F{}^{\mu\nu}=0\,,
\end{align}
where $\nabla^{\mu}$ denotes the covariant derivative associated to our FRW geometry and $\tilde{F}^{\mu\nu}=\frac12\epsilon^{\mu\nu\rho\lambda}F_{\rho\lambda}$ with $\epsilon_{0123}=+\sqrt{|g|}$. We then define $F^{0i}=a^{-2}\,E^i$, $F^{ij}=a^{-2}\,\epsilon_{ijk}\,B^k$, with $\epsilon_{123}=\epsilon^{123}=+1$. In terms of the electromagnetic fields ${\bf E}$ and ${\bf B}$ the equations of motion take the form
\begin{align}
&\partial_\tau^2\Phi+2\,\frac{\partial_\tau a}{a}\,\partial_\tau\Phi-\Delta\,\Phi+a^{2}\frac{dV(\Phi)}{d\Phi}=\frac{\alpha}{f}\,a^{2}\,{\bf E} \cdot {\bf B}\,,\nonumber\\
&\partial_\tau(a^{2}\,{\bf E})-{\pmb\nabla}\times(a^{2}\,{\bf  B})=-\frac{\alpha}{f}\,\partial_\tau\Phi\,(a^{2}\,{\bf B})-\frac{\alpha}{f}\,({\pmb\nabla}\Phi)\times(a^{2}\,{\bf E})\,,\nonumber\\
&{\pmb\nabla} \cdot{\bf E}=-\frac{\alpha}{f}({\pmb\nabla} \Phi)\cdot {\bf B} \,.
\end{align}
In addition, the Bianchi identities read
\begin{equation}
\partial_\tau(a^{2}\,{\bf B} )+{\pmb\nabla}\times(a^{2}\,{\bf E})=0\,,\qquad {\pmb\nabla}\cdot{\bf B}=0\,,
\end{equation}
and are identically satisfied if we define the electrostatic potential $A^0$ and the vector potential ${\bf A}$ through $a^2\,{\bf B}\equiv{\pmb\nabla}\times {\bf A}$, $a^2\,{\bf E}=-\partial_\tau{\bf A}-{\pmb\nabla} A^0$. We choose the gauge ${\pmb\nabla}\cdot{\bf  A}=0$, so that the Maxwell equations read
\begin{align}\label{eq:avec_a0}
&\partial_\tau^2{\bf A}+{\pmb\nabla} \,\partial_\tau(A^0)-\Delta\,{\bf A}=\frac{\alpha}{f}\,\partial_\tau\Phi\,{\pmb\nabla}\times {\bf A}-\frac{\alpha}{f}\,{\pmb\nabla}\Phi\times \partial_\tau{\bf A}-\frac{\alpha}{f}\,({\pmb\nabla}\Phi)\times({\pmb\nabla} A^0)\,,\nonumber\\
&\Delta\,A^0=\frac{\alpha}{f}\,({\pmb\nabla}\Phi)\cdot({\pmb\nabla}\times{\bf A})\,.
\end{align}

The last equation is an elliptic partial differential equation, so its solution is unique, and we denote it as
\begin{align}\label{eq:a0sol}
A^0=\frac{\alpha}{f}\,\frac{1}{\Delta}\left({\bf A}\times {\pmb\nabla} \Phi\right)\,,
\end{align}
where 
\begin{align}
\frac{1}{\Delta}f(\bx)\equiv\int\frac{d\by}{4\pi}\,\frac{f(\by)}{|\bx-\by|}\,.
\end{align}

Inserting the solution (\ref{eq:a0sol}) into the first of eqs.~(\ref{eq:avec_a0}) we obtain the equation
\begin{align}\label{eq:a_total}
&\partial_\tau^2{\bf A}+\frac{\alpha}{f}\,\frac{1}{\Delta}\,\left\{{\pmb\nabla} \,\partial_\tau\left[({\pmb\nabla}\Phi)\cdot({\pmb\nabla}\times{\bf A})\right]\right\}-\Delta\,{\bf A}=\frac{\alpha}{f}\,\partial_\tau\Phi\,{\pmb\nabla}\times {\bf A}-\frac{\alpha}{f}\,{\pmb\nabla}\Phi\times \partial_\tau{\bf A}\nonumber\\
&-\frac{\alpha^2}{f^2}\,({\pmb\nabla}\Phi)\times\left\{\frac{1}{\Delta}\Big[{\pmb\nabla} \left[({\pmb\nabla}\Phi)\cdot({\pmb\nabla}\times{\bf A})\right]\Big]\right\}\,,
\end{align}
and one can check with a direct calculation that this equation is consistent with the gauge choice ${\pmb\nabla}\cdot{\bf A}=0$.

The equation for $\Phi$ reads
\begin{equation}\label{eq:totphi}
\partial_\tau^2\Phi+2\frac{\partial_\tau a}{a}\,\partial_\tau\Phi-\Delta\Phi+a^{2}\frac{dV(\Phi)}{d\Phi}=-\frac{\alpha}{f\,a^{2}}\left[\partial_\tau{\bf A}\cdot({\pmb\nabla}\times{\bf A})+\frac{\alpha}{f}\,({\bf A}\times {\pmb\nabla} \Phi)\cdot({\pmb\nabla}\times{\bf A})\right].
\end{equation}

It will prove convenient to work in cosmic time $t$, defined through $dt=a\,d\tau$, in which case the equations of motion read
\begin{align}\label{eq:tot_t}
&\ddot\Phi+3\,\frac{\dot{a}}{a}\,\dot\Phi-\frac{\Delta}{a^2}\Phi+\frac{d\,V(\Phi)}{d\Phi}=-\frac{\alpha}{f\,a^3}\left[\dot{\bf A}\cdot({\pmb\nabla}\times{\bf A})+\frac{\alpha}{a\,f}\,({\bf A}\times {\pmb\nabla} \Phi)\cdot({\pmb\nabla}\times{\bf A})\right]\,,\nonumber\\
&\ddot{\bf A}+\frac{\dot{a}}{a}\,\dot{\bf A}+\frac{\alpha}{a\,f}\,\frac{1}{\Delta}\,\left\{{\pmb\nabla} \frac{d}{dt}\left[({\pmb\nabla}\Phi)\cdot({\pmb\nabla}\times{\bf A})\right]\right\}-\frac{\Delta}{a^2}\,{\bf A}\nonumber\\
&\qquad=\frac{\alpha}{a\,f}\,\dot\Phi\,{\pmb\nabla}\times {\bf A}-\frac{\alpha}{a\,f}\,{\pmb\nabla}\Phi\times \dot{\bf A}-\frac{\alpha^2}{a^2\,f^2}\,({\pmb\nabla}\Phi)\times\left\{\frac{1}{\Delta}\Big[{\pmb\nabla} \left[({\pmb\nabla}\Phi)\cdot({\pmb\nabla}\times{\bf A})\right]\Big]\right\}\,,
\end{align}
where as usual an overdot denotes a derivative with respect to $t$.

%%%%%%%%%%%%%%%%%%%%%%%%%%%%%%%%%%%%%%%%%%%%%%%%%%%%%%%%%%%%%%%%%%%%%%%%%%%
\section{Linearizing the equations around the instant backreaction regime}%
\label{sec:solutions}%%%%%%%%%%%%%%%%%%%%%%%%%%%%%%%%%%%%%%%%%%%%%%%%%%%%%%
%%%%%%%%%%%%%%%%%%%%%%%%%%%%%%%%%%%%%%%%%%%%%%%%%%%%%%%%%%%%%%%%%%%%%%%%%%%

In the previous section we have presented the {\it exact} (except the assumption of a FRW background) equations governing our system. The fields appearing there should be considered as quantum operators. In this section we will consider the field $\Phi$ as a classical expectation value while ${\bf A}$ will still be treated as an operator.  After reviewing the exact solution for the mode functions of the photon that were first found in~\cite{Anber:2006xt} under some simplifying assumptions on the background dynamics, in Subsection~\ref{subsec:beyond_ibr} we perturb around those solutions.

%%%
\subsection{The IBR regime}
\label{subsec:ibr}
%%%

The system of equations~(\ref{eq:tot_t}) can be solved analytically in the case in which the inflaton is homogeneous and both $\dot\Phi\equiv \dot{\bar\Phi}$ and $\dot{a}/a\equiv\bar{H}$ are constant. We assume without loss of generality that $\dot{\bar\Phi}>0$. As discussed in the Introduction, we refer to this regime as the ``instant backreaction (IBR) regime''. We denote with an overbar the quantities in this regime, so that
\begin{align}
    \Phi(t,\,\bx)=\bar\Phi(t_0)+(t-t_0)\,\dot{\bar\Phi}\,,\qquad \dot{a}(t)/a(t)=\bar{H}\,,\qquad {\rm IBR\ regime}
\end{align}
and the quantities $\dot{\bar{\Phi}}$ and $\bar{\bf A}$ satisfy the equations
\begin{align}\label{eq:tot_tback}
   &3\,\bar{H}\,\dot{\bar\Phi}+\frac{d\,V(\Phi)}{d\Phi}\Bigg|_{\Phi=\Phi(t_0)}=-\frac{\alpha}{f}\,e^{-3\,\bar{H}\,t}\left\langle\dot{\bar{\bf{A}}}\cdot({\pmb\nabla}\times\bar{\bf{A}})\right\rangle\,,\nonumber\\
   &{\ddot{\bar{\bf{A}}}}+\bar{H}\,{\dot{\bar{\bf A}}}-e^{-2\bar{H}t}\,\Delta\,{\bar{\bf {A}}}=\frac{\alpha}{f}\,\,e^{-\bar{H}\,t}\,\dot{\bar\Phi}\,\,{\pmb\nabla}\times {\bar{\bf A}}\,.
\end{align}

It is important to note that this is a consistent set of equation only if we assume $\bar{H}$ to be a constant independent on the matter content of the model (which is not the case, as $\bar{H}$ is determined by the Friedmann's equation). In fact, the Hubble parameter, if expressed in terms of $\Phi$ and ${\bf A}$, cannot be constant under the assumption of constant $\dot{\bar\Phi}\neq 0$ unless we assume $\frac{d\,V(\Phi)}{d\Phi}$ to vanish. We will deal with this issue in Subsection~\ref{subsec:beyond_ibr}.

To simplify our notation we introduce the quantities $\xi$, $\tilde\epsilon$, $\epsilon$ and $\eta$ through
\begin{align}
    &\xi\equiv\frac{\alpha\,\dot{\bar\Phi}}{2\,f\,\bar{H}}\,,\qquad \tilde\epsilon \equiv\frac{\dot{\bar\Phi}^2}{2\,\bar{H}^2\,M_P^2}\,\qquad \left(\Longrightarrow \xi=\frac{\alpha\,\sqrt{2\tilde\epsilon}\,M_P}{2\,f}\right)\,,\nonumber\\ &\frac{d\,V(\Phi)}{d\Phi}\Bigg|_{\Phi=\Phi(t_0)}=-3\,\sqrt{2\,\epsilon}\,\bar{H}^2\,M_P\,,\qquad\frac{d^2\,V(\Phi)}{d\Phi^2}\Bigg|_{\Phi=\Phi(t_0)}=3\,\eta\,\bar{H}^2\,,
\end{align}
where we will trade $\dot{\bar\Phi}$ sometimes for $\xi$ to make contact with the existing literature and sometimes for ${\tilde\epsilon}$ to compare with the slow-roll regime, since particle production slows down the rolling of the inflaton compared to the slow-roll case, so that $\tilde\epsilon\le\epsilon$, with the inequality saturated in the absence of particle production.

The second of eqs.~(\ref{eq:tot_t}) shows that photons of a given helicity get exponentially amplified for $\xi\gtrsim 1$. For our choice of signs, the positive helicity mode is amplified, so that the solution can written as
\begin{align}
\bar{\bf A}(t,\,\bx)=\int\frac{d\bk}{(2\pi)^{3/2}}e^{i\bk\cdot\bx}{\boldsymbol{\epsilon}}_+(\hat\bk)\left[\bar{A}_+(k,\,t)\,\hat{a}_+(\bk)+\bar{A}_+^*(k,\,t)\,\hat{a}^\dagger_+(-\bk)\right]\,,
\end{align}
where the positive helicity projector ${\boldsymbol{\epsilon}}_+(\hat\bk)$, normalized to ${\boldsymbol{\epsilon}}_+(\hat\bk)\cdot{\boldsymbol{\epsilon}}_+(-\hat\bk)=1$, satisfies ${\boldsymbol{\epsilon}}^*_+(\hat\bk)={\boldsymbol{\epsilon}}_+(-\hat\bk)={\boldsymbol{\epsilon}}_-(\hat\bk)$ and $i\,\hat{\bk}\times {\boldsymbol{\epsilon}}_+(\hat\bk)={\boldsymbol{\epsilon}}_+(\hat\bk)$. In the IBR regime the mode functions $\bar{A}(k,\,t)$ can be expressed exactly in terms of Whittaker functions, and if $\xi\gtrsim 1$, which is the case we will be interested in, they simplify to 
\begin{align}\label{eq:Aapprox}
    \bar{A}_+(k,\,t)\simeq \frac{e^{\pi\,\xi}}{\sqrt{2k}}\left(\frac{k\,e^{-\bar{H}\,t}}{2\,\bar{H}\,\xi}\right)^{1/4}\,e^{-2\sqrt{2\,\xi\,k\,e^{-\bar{H}\,t}/\bar{H}}}\,.
\end{align}

In the regime where eq.~(\ref{eq:Aapprox}) is valid, the first of eqs.~(\ref{eq:tot_tback}) can be written as
\begin{align}\label{eq:back_epstil-eps}
    &\sqrt{2\,\tilde\epsilon}-\sqrt{2\,\epsilon}=-\frac{\alpha}{3\,f\bar{H}^2\,M_P}\,e^{-3\,H\,t}\,\langle{\dot{\bar{\bf{A}}}}\cdot({\pmb\nabla}\times{\bar{\bf{A}}})\rangle\simeq -\frac{\alpha\,\bar{H}^2}{f\,M_P}\,\frac{7!}{3\times 2^{21}\pi^2\,\xi^4}\,e^{2\pi\xi}\,.
\end{align}

%%%
\subsection{Beyond the IBR regime}
\label{subsec:beyond_ibr}
%%%

As  stated above, the IBR solution cannot be exact. It can however provide, in a regime that we determine below, a good approximate solution for a certain number of efoldings. In this subsection we derive the linearized equations for $\delta\Phi(t)$ and $\delta{\bf A}(\bx,\,t)$, the differences between the ``actual'' solution of eqs.~(\ref{eq:tot_t}) and the IBR solution. Since we will focus on the onset of the strong backreaction regime, when spatial gradients of the inflaton are small, we will neglect the spatial dependence of $\Phi$:  $\Phi(\bx,\,t)\to \Phi(t)$. In summary, we decompose
\begin{align}
    \Phi(\bx,\,t)=\bar\Phi(t_0)+(t-t_0)\,\dot{\bar\Phi}+\delta\Phi(t)\,,
\end{align}
where $t_0$ is some reference time when the IBR solution is supposed to be a good approximation, so that we will set $\delta\Phi(t_0)=\delta\dot\Phi(t_0)=0$. We will discuss the optimal choice of $t_0$ later on, in Section~\ref{sec:num}.

We then introduce a dimensionless time $x=\bar{H}(t-t_0)$. It will also prove useful to rescale $\delta\Phi(t)=2^{3/2}\,\epsilon\,\sqrt{\tilde\epsilon}\,M_P\,\delta\phi(x)$. The first correction to the Hubble parameter $\delta H(t)=H(t)-\bar{H}$ is then given by 
\begin{align}
    \delta H(t)&\simeq \bar{H}\left(-\sqrt{\epsilon\,\tilde{\epsilon}}\,x+\frac23\,\tilde{\epsilon}\,\epsilon\,\delta\phi'-2\,\sqrt{\tilde\epsilon}\,\epsilon^{3/2}\,\delta\phi\right)\,,
\end{align}
where a prime refers to a derivative with respect to $x$.

As a consequence, also the scale factor receives a correction which, to leading order, gives
\begin{align}
    a(t)\simeq a(t_0)\,e^{x}\left(1-\frac{\sqrt{\epsilon\,\tilde{\epsilon}}}{2}\,x^2+\frac23\,\tilde{\epsilon}\,\epsilon\,\delta\phi-2\,\sqrt{\tilde\epsilon}\,\epsilon^{3/2}\,\int_{0}^x\,\delta\phi(x')\,dx'\right)\,.
\end{align}

After taking the expectation value, the leading order component of the first of eqs.~(\ref{eq:tot_t}) then takes the form
\begin{align}\label{eq:deltaphi_complete}
&\delta\phi''+(3+\tilde{\epsilon})\,\delta\phi'+3\,(\eta-2\sqrt{\epsilon\,\tilde\epsilon}+\tilde\epsilon)\,\delta\phi+9\,\epsilon\left(1-\sqrt{\frac{\tilde\epsilon}{\epsilon}}\right)\int_{0}^x\,\delta\phi(x')\,dx'\nonumber\\
&=-\left[\frac32\left(\frac{\eta}{\epsilon}-\sqrt{\frac{\tilde\epsilon}{\epsilon}}\right)\,x-\frac94\left(1-\sqrt{\frac{\tilde\epsilon}{\epsilon}}\right)\,x^2\right]\nonumber\\
&\qquad\qquad\qquad\qquad\qquad\qquad-\frac{\xi\,e^{-3x}}{2\,\tilde\epsilon\,\epsilon\,a(t_0)^2\,M_P^2}\,\left\langle\delta{\bf A}'\cdot(\tilde{\pmb\nabla}\times{\bar{\bf A}})+{\bar{\bf A}}'\cdot(\tilde{\pmb\nabla}\times\delta{\bf A})\right\rangle\,,
\end{align}
where we have used the background equation~(\ref{eq:back_epstil-eps}) and defined the dimensionless derivatives
\begin{align}
\tilde{\pmb\nabla}\equiv\frac{{\pmb\nabla}}{a(t_0)\,\bar{H}}\,,\qquad\tilde\Delta\equiv\frac{\Delta}{a(t_0)^2\,\bar{H}^2}\,.
\end{align}

Eq.~(\ref{eq:deltaphi_complete}) shows why we rescaled the field $\delta\Phi$ to $\delta\phi$: assuming that the slow-roll parameters $\epsilon$ and $\eta$ are of the same order, as is the case for instance in monomial models of inflation, the inhomogeneous terms in that equation, which act as sources for $\delta\phi$, are of the order of unity as long as $x={\cal O}(1)$. As a consequence, we expect $\delta\phi$ to be naturally of the order of unity for ${\cal O}(1)$ efoldings.

Proceeding in an analogous way, we can write the equation for the first order perturbation of the gauge field as
\begin{align}\label{eq:vec_deltaa}
    &\delta{\bf A}''+\delta{\bf A}'-e^{-2x}\,\tilde\Delta\,\delta{\bf A}-2\,\xi\,\,e^{-x}\,\tilde{\pmb\nabla}\times \delta{\bf A}=-\left(-\sqrt{\epsilon\,\tilde{\epsilon}}\,x+\frac23\,\tilde{\epsilon}\,\epsilon\,\delta\phi'-2\,\sqrt{\tilde\epsilon}\,\epsilon^{3/2}\,\delta\phi\right){\bar{\bf {A}}}'\nonumber\\
    &+e^{-2x}\, \left(\sqrt{\epsilon\,\tilde{\epsilon}}\,x^2-\frac43\,\tilde{\epsilon}\,\epsilon\,\delta\phi+4\,\sqrt{\tilde\epsilon}\,\epsilon^{3/2}\,\int_{0}^x\,\delta\phi(x')\,dx'\right)\,\tilde\Delta\,{\bar{\bf A}}\nonumber\\
    &+2\,\xi\,e^{-x}\,\left(\frac{\sqrt{\epsilon\,\tilde{\epsilon}}}{2}\,x^2-\frac23\,\tilde{\epsilon}\,\epsilon\,\delta\phi+2\,\sqrt{\tilde\epsilon}\,\epsilon^{3/2}\,\int_{0}^x\,\delta\phi(x')\,dx'+2\,\epsilon\,\delta\phi'\right)\,\tilde{\pmb\nabla}\times {\bar{\bf A}}\,.
\end{align}

To write this as an equation for functions rather than operators we decompose the gauge field into mode functions
\begin{align}\label{eq:decomp_a}
{\bf A}(\bx,\,t)=\sum_{\lambda=\pm1}\int \frac{d\bk}{(2\pi)^{3/2}}\,{\boldsymbol{\epsilon}}_\lambda(\hat\bk)\,e^{i\bk\bx}\,\left[A_\lambda(k,\,t)\,\hat{a}_\lambda(\bk)+A^*_\lambda(k,\,t)\,\hat{a}_\lambda(-\bk)\right]\,.
\end{align}

Since eq.~(\ref{eq:vec_deltaa}) does not couple modes of different helicities, and the excited component of $\bar{\bf A}$ contains only positive helicity modes, only the positive helicity of $\delta {\bf A}$ will be excited. As a consequence, we will ignore $A_-(k,\,t)$ altogether in eq.~(\ref{eq:decomp_a}), setting
\begin{align}
    A_+(k,\,t)=\bar{A}_+(k,\,t)+\delta A_+(k,\,t)\,,
\end{align}
where in the regime of large-ish $\xi$ we are considering, $\bar{A}_+(k,\,t)$ is given by eq.~(\ref{eq:Aapprox}).

Inserting the decompositions above into eq.~(\ref{eq:vec_deltaa}) we obtain an equation where $\delta A_+$ is sourced by a term proportional to $\bar{A}_+$. This can be solved using the retarded propagator $D_{\tilde{k}}(x,\,x')$ for the operator
\begin{align}
    \frac{d^2}{dx^2}+\frac{d}{dx}+\tilde{k}^2\,e^{-2x}-2\,\xi\,\tilde{k}\,e^{-x}\,.
\end{align}
Since this is the same operator that gives zero when acting on $\bar{A}_+ (k=\tilde{k}\,a(t_0)\,\bar{H},\,t=t_0+x/\bar{H})\equiv \bar{B}_+(\tilde{k},\,x)$, we have 
\begin{align}\label{eq:propag_general}
    D_{\tilde{k}}(x,\,x')=\frac{\bar{B}_+(\tilde{k},\,x)\,\bar{B}_+(\tilde{k},\,x')^*-\bar{B}_+(\tilde{k},\,x')\,\bar{B}_+(\tilde{k},\,x)^*}{\bar{B}_+'(\tilde{k},\,x')\,\bar{B}_+(\tilde{k},\,x')^*-\bar{B}_+(\tilde{k},\,x')\,\bar{B}'_+(\tilde{k},\,x')^*}\,\Theta(x-x')\,,
\end{align}
(where $\Theta$ denotes the Heaviside step function), which, in the regime $\xi\gtrsim 1$, simplifies to\footnote{Note that one cannot obtain eq.~(\ref{eq:propag_approx}) by simply taking the large-$\xi$ limit~(\ref{eq:Aapprox}) of $\bar{A}_+$ and plugging into eq.~(\ref{eq:propag_general}). In fact, in eq.~(\ref{eq:Aapprox}) we have kept only the growing mode of $\bar{A}_+$, so that the expression in eq.~(\ref{eq:Aapprox}) and its complex conjugate are not linearly independent. To obtain eq.~(\ref{eq:propag_approx}) one has to keep also the decaying mode of $\bar{A}_+$, see eq.~(2.6) in~\cite{Peloso:2022ovc}.}
\begin{align}\label{eq:propag_approx}
D_{\tilde{k}}(x,\,x')=-\frac{e^{-x/4+3\,x'/4}}{\sqrt{2\,\xi\,\tilde{k}}}\sinh\left[2\sqrt{2\,\xi\,{\tilde{k}}}\left(e^{-x/2}-e^{-x'/2}\right)\right]\,\Theta(x-x')\,.
\end{align}

Our final result is 
\begin{align}\label{eq:deltaaplus_fin}
&\delta A_+(\tilde{k},\,x)=\sqrt{\epsilon\,\tilde{\epsilon}}\,e^{\pi\xi}\int_0^x dx'\,\frac{D_{\tilde{k}}(x,\,x')}{\sqrt{2\,\tilde{k}\,a(t_0)\,\bar{H}}}\,\left[\left(x'+2\,\epsilon\,\delta\phi(x')-\frac23\,\sqrt{\epsilon\,\tilde{\epsilon}}\,\frac{d\,\delta\phi}{dx'}\right)\,\sqrt{2\,\xi\,\tilde{k}\,e^{-x'}}\right.\nonumber\\
&+4\,\xi\,\tilde{k}\,e^{-x'}\left(\sqrt{\frac{\epsilon}{\tilde\epsilon}}\,\frac{d\,\delta\phi}{dx'}-\frac{\sqrt{\tilde{\epsilon}\,\epsilon}}{3}\,\delta\phi(x')+\frac{x'{}^2}{4}+\epsilon\int_{0}^{x'} \phi_1(x'')\,dx''\right)\nonumber\\
&+\left.4\,k^2\,e^{-2\,x'}\left(\frac{\sqrt{\epsilon\,\tilde{\epsilon}}}{3}\,\delta\phi(x')-\frac{x'{}^2}{4}-\epsilon\,\int_{0}^{x'}\,\delta\phi(x'')\,dx''\right)\right]\left(\frac{e^{-x'}\,\tilde{k}}{2\,\xi}\right)^{1/4}\,e^{-2\sqrt{2\,\xi\,\tilde{k}\,e^{-x'}}}\,,
\end{align}
where we have set $\delta A_+(\tilde{k},\,x=0)=0$ as we assume that we start evolving the system in the weak backreaction regime when the IBR solution is still valid.

Before trying to solve the system of equations (\ref{eq:deltaphi_complete}) and (\ref{eq:deltaaplus_fin}) we note that the expectation value appearing on the right hand side of eq.~(\ref{eq:deltaphi_complete}) can be written in terms of the functions $\bar{A}_+$ and $\delta A_+$ as
\begin{align}\label{eq:exp_adeltaa}
\left\langle\delta{\bf A}'\cdot(\tilde{\pmb\nabla}\times{\bar{\bf A}})+{\bar{\bf A}}'\cdot(\tilde{\pmb\nabla}\times\delta{\bf A})\right\rangle=-a(\bar{t})^3\,\bar{H}^3\int\frac{\tilde{k}^3\,d\tilde{k}}{2\pi^{2}}\,\frac{d}{dx}\left(\bar{A}_+\,\delta A_+\right)\,.
\end{align}

%%%%%%%%%%%%%%%%%%%%%%%%%%%%%%%%%%%%%%%%%%%%%%%%%%%%%%%%%%%%%%%%
\section{Solving the linearized system, including source terms}%
\label{sec:lap}%%%%%%%%%%%%%%%%%%%%%%%%%%%%%%%%%%%%%%%%%%%%%%%%%
%%%%%%%%%%%%%%%%%%%%%%%%%%%%%%%%%%%%%%%%%%%%%%%%%%%%%%%%%%%%%%%%

We are now in position to collect the equations derived in the previous section and solve them. To do so, we first use the expressions~(\ref{eq:Aapprox}) and~(\ref{eq:deltaaplus_fin}) for $\bar{A}_+$ and $\delta A_+$ respectively, where the propagator is given by eq.~(\ref{eq:propag_approx}), and insert them into eq.~(\ref{eq:exp_adeltaa}). Then we insert the resulting expression into eq.~(\ref{eq:deltaphi_complete}).  This gives an integro-differential equation for $\delta\phi(x)$ only. We simplify this equation by {\it (i)} assuming a quadratic inflaton potential (i.e., $\eta=\epsilon$) which, even if ruled out by observations, is often used as a benchmark in the literature, {\it (ii)} keeping only the leading terms in the large-$\xi$ expansion, and {\it (iii)} keeping only the leading order terms in the slow-roll expansion, assuming $\tilde\epsilon$ to be of the same order of $\epsilon$.

We then introduce the parameter 
\begin{align}\label{eq:defkappa}
\kappa\equiv\sqrt{\frac{\ep}{\tilde{\ep}}}-1\,,
\end{align}
which, since it vanishes in the absence of particle creation, is a measure of the strength of the backreaction. Since $\tilde\epsilon\le\epsilon$, $\kappa\ge 0$. Using the background equation~(\ref{eq:back_epstil-eps}), we can write 
\begin{align}\label{eq:alsokappa}
    \kappa&=\frac{105}{2^{17}}\frac{\bar{H}^2}{\tilde{\ep}\,M_P^2}\frac{e^{2\pi\xi}}{\pi^2\,\xi^3}=\left|\frac{\alpha\langle{\bf E}\cdot{\bf B}\rangle/f}{3\,\bar{H}\,\dot{\bar\Phi}}\right|\,.
\end{align}

Once one keeps only the leading terms in the slow-roll approximation, the equation for $\delta\phi$ contains only $\delta\phi''$ and $\delta\phi'$. As a consequence, we define a new variable 
\begin{align}
u(x)\equiv \delta\phi'(x)\,.
\end{align}

After all these manipulations, we finally obtain the master equation whose solution will allow us to determine $\Phi(t)$ to first order in the deviation from the IBR solution,
\begin{align}
\label{eqn:u}
&\frac{d\,u(x)}{dx}+3\,u(x)+\frac32\,\kappa\,x\left(1-3\,x\right)=-\frac{128}{105}\,\kappa\int_0^{4\xi^2}\,dq\,q^{5/2}\,e^{7(x'-x)/2}\nonumber\\
&\int_0^{x}\,dx'\,\left[\Big(e^{-4\sqrt{q}}-e^{-4\sqrt{q}\,e^{(x'-x)/2}}\Big)+4\,\sqrt{q}\,e^{(x'-x)/2}e^{-4\sqrt{q}\,e^{(x'-x)/2}}\right]
\left[x'+\frac{\sqrt{q}}{2}\left(x'{}^2+4\,u(x')\right)\right]\,
\end{align}
where we have redefined the dimensionless momentum $\tilde{k}\equiv q\,e^{x'}/(2\,\xi)$.

Before we proceed we must discuss the upper integration limit for the integral in $dq$ in eq.~(\ref{eqn:u}). In principle this limit is not needed, as that integral is UV-convergent. In ref.~\cite{Peloso:2022ovc}, however, a UV cutoff was made necessary by the fact that the integration region in $dx'$ (or more precisely, its equivalent integral over conformal time) was extending all the way to $x'\to-\infty$. In the absence of a UV cutoff in the integral in $dq$, the subsequent integral in $dx'$ would diverge at $x'\to-\infty$. Such a cutoff, on the other hand, is not required in our case, since the integral in $dx'$, which extends from $x'=0$ to $x'=x$, is now convergent even if the integral in $dq$ is extended all the way to $q\to\infty$. However, by removing the UV cutoff from the integral in $dq$, we would obtain results that are qualitatively inconsistent with those obtained in the literature, as we show in Appendix~\ref{app:cutoff}. As we will see below, introducing a hard UV cutoff for that integral allows us to instead reproduce the results obtained by full numerical studies with very good accuracy. For this reason, we will follow~\cite{Peloso:2022ovc} and impose a hard cutoff on the integral in $dq$. We will place the cutoff at $q=4\,\xi^2$, which corresponds the value of $q$ at which the gauge field mode functions start feeling a tachyonic instability.

Equation~(\ref{eqn:u}) is a linear, inhomogeneous, integro-differential equation. Even if it looks quite a formidable equation, we can solve it using semi-analytical methods (the only numerical step being the search of the poles of an holomorphic function), as we now discuss.

We take the Laplace transform of both sides of eq.~\eqref{eqn:u}, which allows us to write the source term as a sum of convolutions. In Appendix~\ref{app:Laplace} we give a detailed explanation on how this is done. Defining the Laplace transform operator as $\mathcal{L}[\,]$, $\mathcal{L}[u](p)=U(p)$, we obtain
\begin{align}
p\,U(p)-u(0)+3\,U(p)+\frac{3}{2\,p^2}\,\kappa\left(1-\frac{6}{p}\right)
=
-\frac{2^7}{105}\,\kappa
\left[
\mathcal{K}_1(p)\left(\frac{1}{p^3}+2\,U(p)\right)+\frac{\mathcal{K}_2(p)}{p^2}\right],
\end{align}
where for $\xi\gg 1$ we have
\begin{align}
    \mathcal{K}_1(p)&\simeq \frac{p+3}{(2p-1)(2p+7)}\Bigg[\frac{315}{64}-\frac{\Gamma(8+2p)}{2^{10+6p}\,\xi^{2p-1}}\Bigg],\quad \mathcal{K}_2(p)\simeq \frac{p+3}{p\,(2p+7)}\Bigg[\frac{315}{256}-\frac{\Gamma(8+2p)}{2^{19+6p}\,\xi^{2p}}\Bigg].
\end{align}
Now we see the power of the Laplace transform, which has allowed to convert a scary integro-differential equation into a simple equation for $U(p)$, which we solve algebraically, obtaining 
\begin{align}\label{eq:ufinalshort}
U(p)&=\frac{\mathcal{N}(p)}{\mathcal{D}(p)},
\end{align}
where we have set $u(0)=0$ and
\begin{align}\label{eq:nd}
\mathcal{N}(p)
&=
\kappa\,\left(-\frac{3p}{2}+9\right)(2p-1)(2p+7)
\nonumber\\
&\quad
-\frac{2}{105}(p+3)\,\kappa\,
\Bigg[
315
+\frac{315}{4}(2p-1)
-\frac{\Gamma(8+2p)}{2^{4+6p}\xi^{2p-1}}
-\frac{(2p-1)\Gamma(8+2p)}{2^{6+6p}\xi^{2p}}
\Bigg],\nonumber\\
\mathcal{D}(p)
&=
p^3(p+3)\,\mathcal{F}(p)\,,\qquad  \mathcal{F}(p)=
(2p-1)(2p+7)
+
\frac{2^9}{105}\kappa
\left(
\frac{315}{128}
-
\frac{\Gamma(8+2p)}{2^{11+6p}\xi^{2p-1}}
\right)\,.
\end{align}

%%%%%%%%%%%%%%%%%%%%%%%%%%%%%%
%%%%%%%%%%%%%%%%%%%%%%%%%%%%%%
%%%%%%%%%%%%%%%%%%%%%%%%%%%%%%
%%%%%%%%%%%%%%%%%%%%%%%%%%%%%%
\begin{figure}[t]
    \centering

    \begin{subfigure}[t]{0.48\linewidth}
        \centering
        \includegraphics[width=\linewidth]{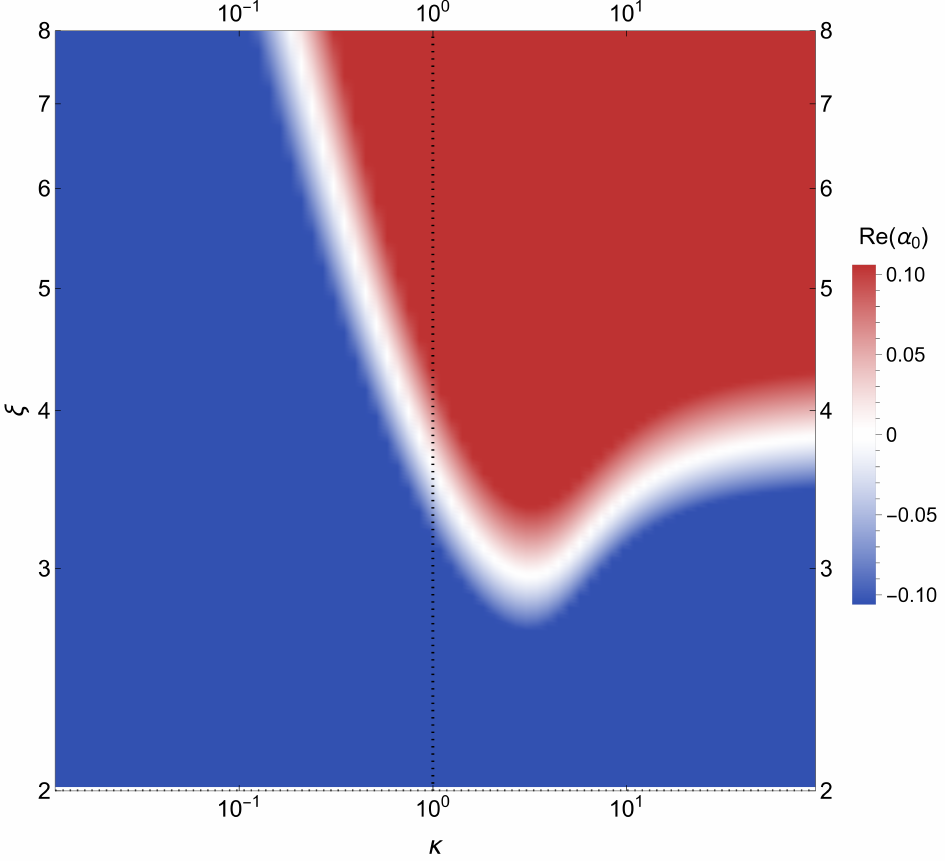}
        \caption{}
        \label{fig:densitygrad}
    \end{subfigure}
    \hfill
    \begin{subfigure}[t]{0.43\linewidth}
        \centering
        \includegraphics[width=\linewidth]{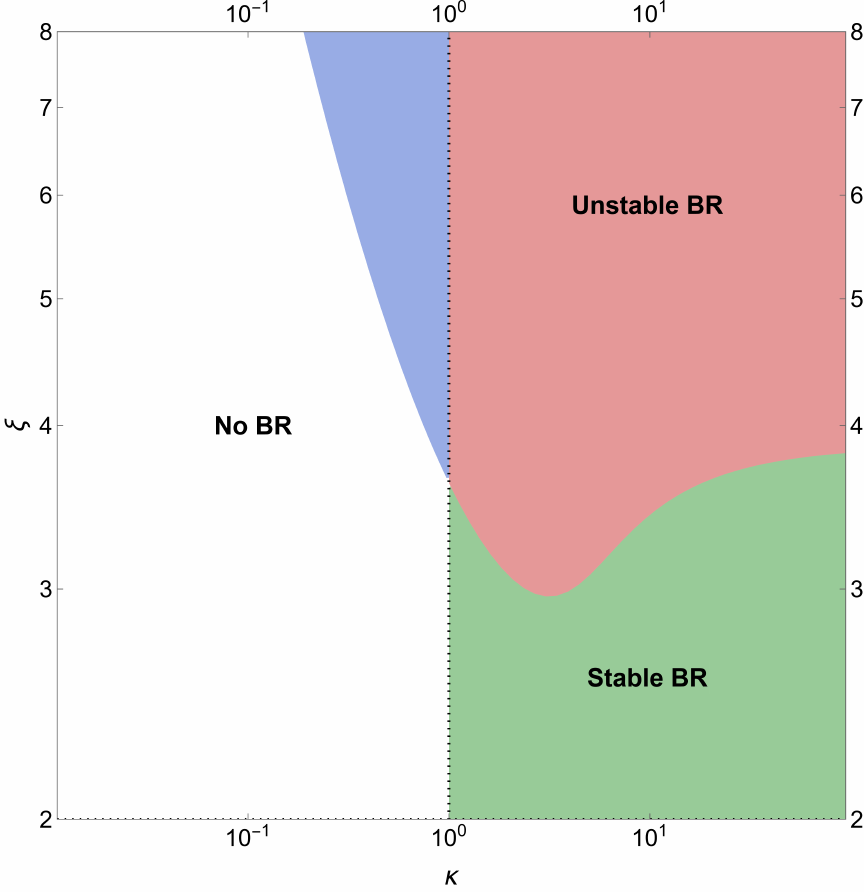}
        \caption{}
        \label{fig:densitycolor}
    \end{subfigure}
    \caption{Density plots of $\mathrm{Re}(\alpha_0)$ in the $(\kappa,\xi)$ plane. (a) The blue region corresponds to $\mathrm{Re}(\alpha_0)<0$, the red region to $\mathrm{Re}(\alpha_0)>0$, and the shaded white band marks the transition between the two. (b) The white region denotes $\mathrm{Re}(\alpha_0)<0$ with $\kappa<1$ (no BR), the blue region $\mathrm{Re}(\alpha_0)>0$ with $\kappa<1$ (unstable weak BR), the red region $\mathrm{Re}(\alpha_0)>0$ with $\kappa>1$ (unstable strong BR), and the green region $\mathrm{Re}(\alpha_0)<0$ with $\kappa>1$ (stable strong BR). The black dotted line indicates $\kappa=1$.}
    \label{fig:densityplots}
\end{figure}
%%%%%%%%%%%%%%%%%%%%%%%%%%%%%%
%%%%%%%%%%%%%%%%%%%%%%%%%%%%%%
%%%%%%%%%%%%%%%%%%%%%%%%%%%%%%
%%%%%%%%%%%%%%%%%%%%%%%%%%%%%%

Finally, to obtain $u(x)$ we must take the inverse Laplace transform of $U(p)$, which is determined by the singularity structure of $U(p)$. Essentially, for each simple pole at $p=p_0$, the inverse Laplace transform contains a term going as $e^{p_0x}$, whose coefficient is determined by the corresponding residue. For multiple poles the inverse Laplace transform contains a polynomial contribution. In particular, the relevant poles are obtained from the zeros of the denominator $\mathcal{D}(p)$. For the range of $\kappa$ and $\xi$ considered here, the pole structure consists of three distinct contributions. 

First, there is a third-order pole at $p=0$, which gives a polynomial contribution in real space,
\begin{align}
u_0(x)=a_0\,x^2+b_0\,x+c_0,
\end{align}
where the coefficients $a_0$, $b_0$ and $c_0$, that we do not write down here, depend on the residues of $U(p)$ at the pole $p=0$.

Second, there is a simple pole at $p=-3$, leading to a decaying mode of the form
\begin{align}
u_{-3}(x)=a_{-3}\,e^{-3\,x}.
\end{align}
Again, the constant $a_{-3}$ depends on the residue of $U(p)$ at $p=-3$ and we do not express it here.

Finally, the remaining poles are determined by the transcendental part of the denominator, which we write, from eq.~(\ref{eq:nd}), as ${\cal F}(p)=0$. These poles appear as complex conjugate pairs and therefore contribute
\begin{align}
u_{\cal F}(x)=2\,\mathrm{Re}\!\,\left[\beta_0\, e^{\alpha_0 \,x}\right]+2\,\sum_{i\geq1} \mathrm{Re}\!\,\left[\beta_i\, e^{\alpha_i \,x}\right],
\end{align}
where $\alpha_0$ is the rightmost (i.e., with the largest real part) complex pole pair  with the residue $\b_0$ and $\alpha_i$, $\beta_i$ are the remaining pole and residue pairs. Depending on the values of $\kappa$ and $\xi$, the rightmost pole $\alpha_0$ will or will not have a positive real part. This mode controls the late-time behavior of the system: ${\mathrm {Re}}(\alpha_0)>0$ signals unstable back reaction. One should note that $p=\frac{1}{2}$ is also a zero of ${\cal F}(p)$ but that contribution is canceled by the fact that $\mathcal{N}(\frac{1}{2})=0$ for any value for the pair $(\kappa,\,\xi)$.

Figure~\ref{fig:densitygrad} shows the parameter space in the $(\kappa,\,\xi)$ plane. The red and blue regions correspond to unstable (${\mathrm {Re}}(\alpha_0)>0$) and stable (${\mathrm {Re}}(\alpha_0)<0$) backreaction respectively. In the standard characterization of backreaction, $\kappa\gtrsim 1$ corresponds to the strong-backreaction regime, while $\kappa\lesssim 1$ corresponds to weak backreaction. 

By solving a single equation ${\cal F}(p)=0$, we recover the regions of parameter space identified in ref.~\cite{Sobol:2026abc}. In particular, as shown in fig.~\ref{fig:densitycolor}, there is an unstable weak-backreaction region, shown in blue, where $\kappa<1$, but $\mathrm{Re}(\alpha_0)>0$. In this regime the gauge field-induced friction is subdominant compared to force acting on the inflaton from the potential gradient, yet the IBR background solution is unstable. Conversely, there is also a stable strong-backreaction region, shown in green, where $\kappa>1$ but $\mathrm{Re}(\alpha_0)<0$. This latter region appears to extend all the way to very large values of $\kappa$, provided $\xi$ is small enough. It is worth stressing that spatial gradients of the inflaton, that have been neglected in this work, might provide an additional channel for an instability, even in regions that are marked as stable in fig.~\ref{fig:densitygrad}.
Note also that all the formulae used in this paper rely on the large-$\xi$ limit of the gauge field mode functions. However, $\xi\gtrsim 3$ is usually sufficient to provide good approximations.  In particular, the region of stable backreaction (green in fig.~\ref{fig:densitycolor}) has $\xi \lesssim 4$, which is probably too small to lead to observable signatures such as gravitational waves at frequencies probed by interferometers. Nevertheless, we consider this region  of great interest as it {\it might} allow to realize a novel regime of smooth slow-roll inflation~\cite{Anber:2009ua}, which could open new avenues for model building, following the philosophy behind, for instance, the model presented in~\cite{Roberts:2021plm}.

%%%%%%%%%%%%%%%%%%%%%%%%%%%%%%%%%%%%%%%%%%%%%
%%%%%%%%%%%%%%%%%%%%%%%%%%%%%%%%%%%%%%%%%%%%%
%%%%%%%%%%%%%%%%%%%%%%%%%%%%%%%%%%%%%%%%%%%%%
%%%%%%%%%%%%%%%%%%%%%%%%%%%%%%%%%%%%%%%%%%%%%
\begin{figure}
    \centering
    \includegraphics[width=0.4\linewidth]{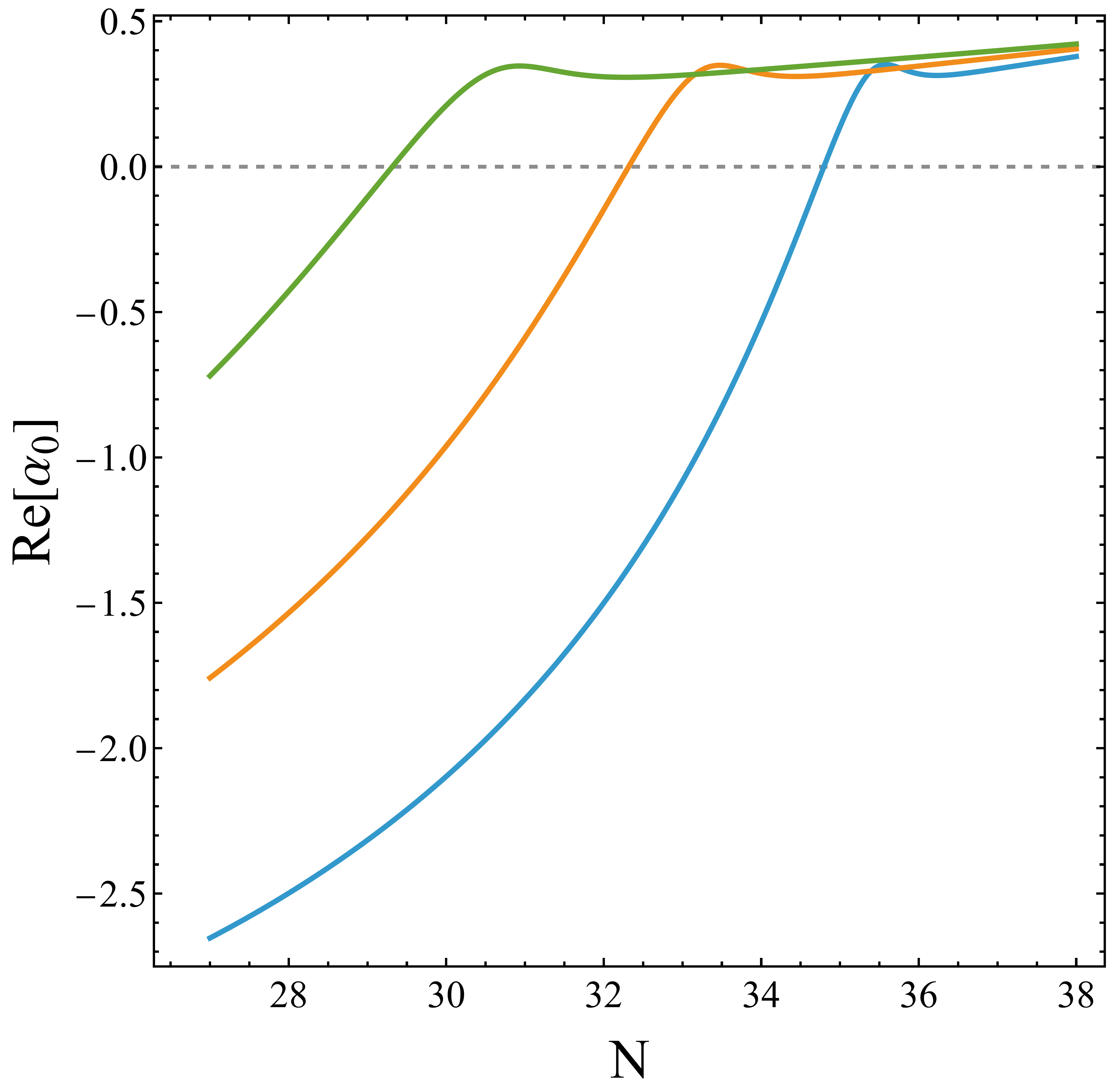}
    \caption{Evolution of $\mathrm{Re}(\alpha_0)$ for $\frac{\alpha\,M_p}{f}=25,\,30,\, 35$, shown (left to right) by the green, orange, and blue curves, respectively. The black dotted line marks $\mathrm{Re}(\alpha_0)=0$; each crossing of this line indicates a change in the stability of the corresponding solution, with $\mathrm{Re}(\alpha_0)>0$ denoting instability and $\mathrm{Re}(\alpha_0)<0$ denoting stability.}
    \label{fig:nbarselect}
\end{figure}
%%%%%%%%%%%%%%%%%%%%%%%%%%%%%%%%%%%%%%%%%%%%%
%%%%%%%%%%%%%%%%%%%%%%%%%%%%%%%%%%%%%%%%%%%%%
%%%%%%%%%%%%%%%%%%%%%%%%%%%%%%%%%%%%%%%%%%%%%
%%%%%%%%%%%%%%%%%%%%%%%%%%%%%%%%%%%%%%%%%%%%%

%%%%%%%%%%%%%%%%%%%%%%%%%%%%%%%%%%%%%%%%%%%%%%%%%%%%%%%%%%%%%%%%%%%%%%%%%%%%%%%%%
\section{The development of the instability in time. Comparison with numerical results}%%
\label{sec:num}%%%%%%%%%%%%%%%%%%%%%%%%%%%%%%%%%%%%%%%%%%%%%%%%%%%%%%%%%%%%%%%%%%
%%%%%%%%%%%%%%%%%%%%%%%%%%%%%%%%%%%%%%%%%%%%%%%%%%%%%%%%%%%%%%%%%%%%%%%%%%%%%%%%%

As discussed above, it is possible to follow the development of the instability in time by solving analytically the equation for $\delta\Phi(t)$, including the source term. In this section we check the validity of our formulae by comparing the evolution of the parameter $\xi$ obtained by solving numerically the full system of (spatially homogeneous) axion$+$gauge field equations with the analytic approximation constructed using the techniques described in the previous sections. We perform this comparison for a quadratic potential, $V(\Phi)=\frac{m^2}{2}\Phi^2$, with $m=6\times 10^{-6}\,M_{\rm P}$, and for three representative values of the coupling, $\frac{\alpha M_{\rm P}}{f}=25,\,30,\,35$. In this section we use as time variable the number of efoldings $N$, which to leading order in slow-roll is given by $dN=\bar{H}\,dt=dx$.

As shown in the previous section, the function $u(x)$ (and, as a consequence, $\delta\Phi(N)$) can be written as a combination of polynomial terms and complex exponentials. Using the first-order expansions of the Hubble parameter and the homogeneous field $\Phi$, the evolution of $\xi$ can therefore be expressed, up to first order in the slow-roll parameters, as the sum of the slow-roll contribution and the correction induced by backreaction,
\begin{align}
\label{eqn:xiapx}
\xi(N)
&=\xi(\bar{N})\,\Bigg[
1+\sqrt{\epsilon\,\tilde{\epsilon}}\,x\\
&\qquad\qquad +2\sqrt{\epsilon\,\tilde{\epsilon}}\,
\Bigg(
a_0 x^2+b_0 x+c_0+a_{-3}e^{-3x}
+2\,\mathrm{Re}\!\left[\beta_0 e^{\alpha_0 x}\right]
+2\sum_{i\geq1}\mathrm{Re}\!\left[\beta_i e^{\alpha_i x}\right]
\Bigg)
\Bigg]\,,\nonumber
\end{align}
where $\bar{N}=N(t_0)$ and $x=N-\bar{N}$.

%%%%%%%%%%%%%%%%%%%%%%%%%%%%%%%
%%%%%%%%%%%%%%%%%%%%%%%%%%%%%%%
%%%%%%%%%%%%%%%%%%%%%%%%%%%%%%%
%%%%%%%%%%%%%%%%%%%%%%%%%%%%%%%
\begin{figure}[t]
    \centering
    \begin{subfigure}[t]{0.32\linewidth}
        \centering
        \includegraphics[width=\linewidth,height=0.8\linewidth]{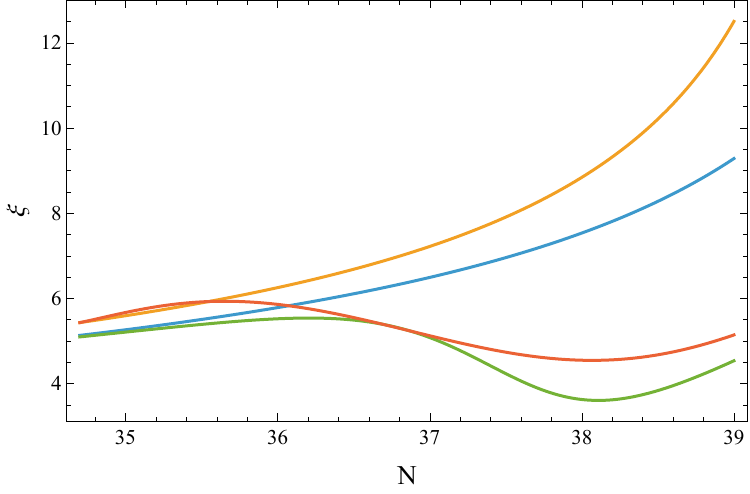}
        \caption{}
        \label{fig:xi25}
    \end{subfigure}
    \hfill
    \begin{subfigure}[t]{0.32\linewidth}
        \centering
        \includegraphics[width=\linewidth,height=0.8\linewidth]{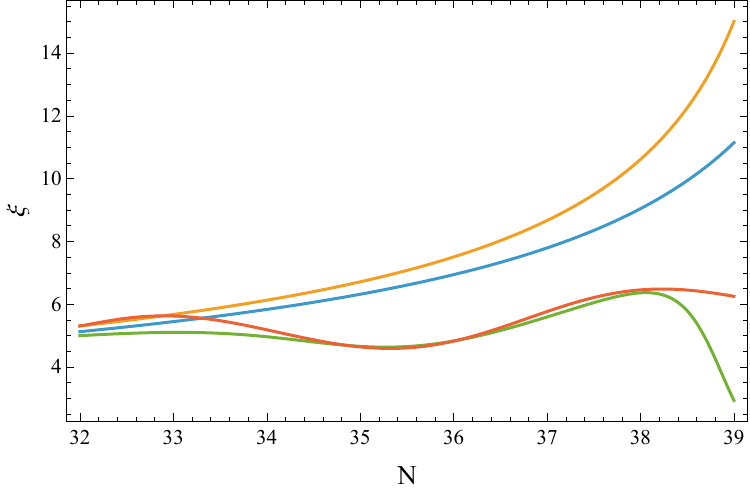}
        \caption{}
        \label{fig:xi30}
    \end{subfigure}
    \hfill
    \begin{subfigure}[t]{0.32\linewidth}
        \centering
        \includegraphics[width=\linewidth,height=0.8\linewidth]{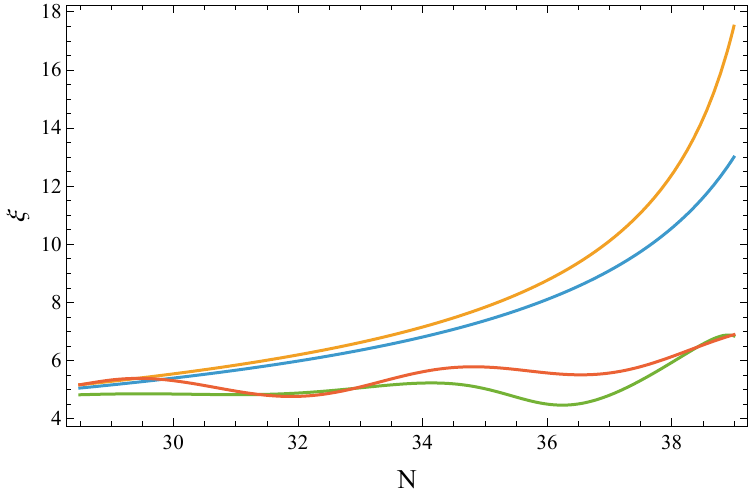}
        \caption{}
        \label{fig:xi35}
    \end{subfigure}
    \caption{Evolution of $\xi$ parameter. From top to bottom in each panel, the orange line is the slow roll evolution, the blue line is the exact evolution in the absence of backreaction, the red line is the first order approximate evolution obtained using our analytical formulae, and the green line is the full numerical solution, including backreaction, for (a): $\frac{\alpha\,M_p}{f}=25$, (b): $\frac{\alpha\,M_p}{f}=30$, (c): $\frac{\alpha\,M_p}{f}=35$. }
    \label{fig:xiplots}
\end{figure}
%%%%%%%%%%%%%%%%%%%%%%%%%%%%%%%
%%%%%%%%%%%%%%%%%%%%%%%%%%%%%%%
%%%%%%%%%%%%%%%%%%%%%%%%%%%%%%%
%%%%%%%%%%%%%%%%%%%%%%%%%%%%%%%

For a given model, once $t_0$ (i.e., $\bar{N}$) is specified, the constants appearing in eq.~\eqref{eqn:xiapx} are determined by the corresponding values of $\xi$ and $\kappa$. But, how to determine $t_0$? It turns out that what is arguably the simplest choice for $t_0$ is also the one that gives the best results: by choosing it to be the earliest time at which the real part of the rightmost pole of $U(p)$ first becomes positive, our analytical solution follows quite closely the numerical evolution.  This criterion identifies the onset of the instability of the stationary background solution. Figure~\ref{fig:nbarselect} shows how $\mathrm{Re}[\alpha_0]$ varies as the expansion point $\bar{N}$ is changed; the selected value of $\bar{N}$ is the point at which this curve crosses zero.

Our numerical study uses the implementation described in Appendix A of~\cite{Garcia-Bellido:2023ser}, where the axion background and gauge field modes are evolved simultaneously on a logarithmic grid of comoving momenta. The backreaction integral is reconstructed from this grid, with a time-dependent UV cutoff that removes vacuum modes and includes only modes that have previously entered the tachyonic regime. We count the number of efoldings $N$ as an increasing function of time, setting $N=40$ at the time at which inflation would end if the slow roll approximation were valid throughout the inflationary phase. Figure~\ref{fig:xiplots} compares the numerical and analytical evolution of $\xi$ for $\frac{\alpha M_p}{f}=25$ (fig.~\ref{fig:xi25}), $\frac{\alpha M_p}{f}=30$ (fig.~\ref{fig:xi30}), and $\frac{\alpha M_p}{f}=35$ (fig.~\ref{fig:xi35}). In particular, in those figures the green curves show the numerical solution of the coupled equations, while the red curve shows the analytical approximation given by Eq.~\eqref{eqn:xiapx}. The agreement between those curves demonstrates that the expansion around the stationary solution captures the leading backreaction correction to the evolution of $\xi$. While this agreement is not perfect, we find it good enough also in view of the fact that it is based on a truncation of our equations to first order in the slow roll approximation.

%%%%%%%%%%%%%%%%%%%%%%%%%%
\section{Summary}%%%%%%%%%
\label{sec:summary}%%%%%%%
%%%%%%%%%%%%%%%%%%%%%%%%%%

There is by now a substantial literature on the strong backreaction regime of axion inflation. To our knowledge, the present work is the first one that provides relatively simple analytical formulae (all one needs to do is to determine the pole structure of the analytical function obtained from eqs.~(\ref{eq:ufinalshort}) and~(\ref{eq:nd})) to determine the time evolution of the system at the transition to the strong backreaction regime. 

This work shares several features with~\cite{Peloso:2022ovc}, the most important of which is that it is based on a perturbative expansion around the IBR solution. Moreover, the equation determining the poles of the Laplace transform is the same as that  used in~\cite{Peloso:2022ovc} to determine the Lyapunov exponents for $\delta\Phi$. In addition, however, the present work contains a full scan of the parameter space of the system (see fig.~\ref{fig:densitycolor}). This scan shows the existence of a regime of stable strong backreaction, consistently with the results recently found with completely different methods in~\cite{Sobol:2026abc}. We have also observed (see Appendix~\ref{app:cutoff}) that the presence of a UV cutoff plays a crucial role in reproducing the results obtained numerically.

Our formulae, which we have tested against the full numerical evaluations for a quadratic potential and for a set of values of the coupling,  can be quickly adapted to arbitrary inflationary potentials and couplings. These formulae can be useful to quickly scan the parameter space of axion inflation for phenomenological applications such as that in~\cite{Garny:2026gcs}.

We see at least two directions in which our formalism could be useful. First, by keeping, to first order, the space-dependent part of $\Phi$ in eqs.~(\ref{eq:tot_t}), it should be possible to derive an analytical expression for the amplification of spatial gradients of the inflaton as the instability develops. The generation of large gradients in the strong backreaction regime has been an interesting (and to some extent, unexpected) outcome of the lattice studies of the system~\cite{Figueroa:2023oxc,Figueroa:2024rkr} (which however is model dependent~\cite{Barbon:2025wjl}). An analytical study, complementary to the very recent analysis of~\cite{Domcke:2026fjo}, might help clarifying their origin. Second, it would be interesting to explore in greater detail the region of stable backreaction, and having explicit analytical formulae would help shed some light on the nature of the system in that region. We hope to come back to these questions in future publications.

%%%%
\acknowledgments 
%%%%

We thank Florian Niedermann for useful discussions, and Valerie Domcke and Marco Peloso for interesting correspondence. In particular, we thank Marco Peloso for pointing out a substantial mistake in the first version of eq.~(\ref{eq:a_total}). This work is partially supported by the US-NSF grant PHY-2112800 and PHY-2412570.

%%%%%%%%%%%%%%%%%%%%%%%%%%%%%%%%%%%%
%%%%%%%%%%%%%%%%%%%%%%%%%%%%%%%%%%%%
%%%%%%%%%%%%%%%%%%%%%%%%%%%%%%%%%%%%
%%%%%%%%%%%%%%%%%%%%%%%%%%%%%%%%%%%%
\appendix%%%%%%%%%%%%%%%%%%%%%%%%%%%
%%%%%%%%%%%%%%%%%%%%%%%%%%%%%%%%%%%%
%%%%%%%%%%%%%%%%%%%%%%%%%%%%%%%%%%%%
%%%%%%%%%%%%%%%%%%%%%%%%%%%%%%%%%%%%
%%%%%%%%%%%%%%%%%%%%%%%%%%%%%%%%%%%%

%%%%%%%%%%%%%%%%%%%%%%%%%%%%%
\section{Laplace transform}%%
\label{app:Laplace}%%%%%%%%%%
%%%%%%%%%%%%%%%%%%%%%%%%%%%%%

In this appendix we show how Eq.~\eqref{eqn:u} can be written in convolution form, and illustrate how the Laplace transform of the source terms is obtained. We define the Laplace transform of a function $\hat f(x)$ as
\begin{align}
    \mathcal{L}[\hat f](p)
    \equiv \int_0^\infty dx\, e^{-px}\hat f(x)
    \equiv f(p)\,.
\end{align}

After introducing the variable $ s=x-x'$, the source term in Eq.~\eqref{eqn:u} can be written as a sum of convolutions proportional to
\begin{align}
\int_0^x ds\,
\left[\frac{1}{2}\,\hat{\mathcal{K}}_1(s)
\left(
\frac{(x-s)^2}{4}+u(x-s)
\right)+\hat{\mathcal{K}}_2(s)\,\frac{x-s}{4}
\right].
\end{align}
Here the kernels are defined by
\begin{align}
\hat{\mathcal{K}}_1(s)
&=
\int^{4\xi^2}dy\,y^{3}e^{-7s/2}
\left[
e^{-4\sqrt{y}}
-e^{-4\sqrt{y}e^{-s/2}}
+4\sqrt{y}\,e^{-s/2}e^{-4\sqrt{y}e^{-s/2}}
\right]\,,\nonumber\\
\hat{\mathcal{K}}_2(s)
&=
\int^{4\xi^2}dy\,y^{5/2}e^{-7s/2}
\left[
e^{-4\sqrt{y}}
-e^{-4\sqrt{y}e^{-s/2}}
+4\sqrt{y}\,e^{-s/2}e^{-4\sqrt{y}e^{-s/2}}
\right]\,.
\end{align}
This form is useful because the Laplace transform of a convolution factorizes. In particular,
\begin{align}
\mathcal{L}\left[
\int_0^x ds\,\hat{\mathcal{K}}_2(s)\,\frac{x-s}{4}
\right]
=\frac{\mathcal{K}_2(p)}{4\,p^2},
\end{align}
while
\begin{align}
\mathcal{L}\left[
\int_0^x ds\,\hat{\mathcal{K}}_1(s)
\left(
\frac{(x-s)^2}{4}+u(x-s)
\right)
\right]
=
\mathcal{K}_1(p)
\left(
\frac{1}{2\,p^3}+U(p)
\right).
\end{align}
Therefore, taking the Laplace transform of Eq.~\eqref{eqn:u} gives
\begin{align}
p\,U(p)-u(0)
+3\,U(p)
+\frac{3\kappa}{2p^2}
\left(1-\frac{3}{p}\right)
=
-\frac{2^9}{105}\kappa
\left[
\frac{\mathcal{K}_2(p)}{4\,p^2}
+\frac{1}{2}\mathcal{K}_1(p)
\left(
\frac{1}{2\,p^3}+U(p)
\right)
\right].
\end{align}
This algebraic equation for $U(p)$ follows directly from the convolution theorem and is the starting point for identifying the pole structure that controls the evolution of $u(x)$.

Next, we show how one take perform the Laplace transformation of one of the terms on the source side of eq.~(\ref{eqn:u}). In particular, here we focus only on the term involving $u(x')$. The analysis of the other terms is analogous. Again taking $x-x'=s$, the Laplace transform of this term can be taken by multiplying it with $\int_0^{\infty}dx\,e^{-p\,x}$. Ignoring any constants in front, this reads
\begin{align}
   \int^{\infty}_0\,dx\, &e^{-p\,x}\int^{4\xi^2}\,dy\,y^{3}\,e^{-7s/2}\nonumber\\
   &\times\int_0^{x}\,ds\,\left[\left(e^{-4\sqrt{y}}-e^{-4\sqrt{y}\,e^{-s/2}}\right)+4\,\sqrt{y}\,e^{-s/2}e^{-4\sqrt{y}\,e^{-s/2}}\right]\times u(x-s)\,,
\end{align}
which we write as
\begin{align}
    \int^{\infty}_0\,dx\, e^{-p\,x}\int_0^{x}\,ds\,\hat{\mathcal{K}}_1(s)\,u(x-s)\,.
\end{align}
The integration region is $0\leq s\leq x\leq \infty$. Changing variables to $t=x-s$, we write our integral as
\begin{align}
    &\int_0^{\infty}ds\,\hat{\mathcal{K}}_1(s)\,e^{-p\,s}\int_0^{\infty}\,dt \,u(t)\,e^{-p\,t}=\mathcal{K}_1(p)\times U(p)\,,
\end{align}
with $\mathcal{L}(\hat{\mathcal{K}_1})[p]=\mathcal{K}_1(p)$ and $\mathcal{L}(u)[p]=U(p)$. 

Next we show how to explicitly compute the Laplace transform of the kernel $\hat{\mathcal{K}}$,
\begin{align}
    \mathcal{K}_1(p)&=\int_0^{\infty} ds\,e^{-p\,s}\int^{4\xi^2}\,dy\,y^{3}\,e^{-7s/2}\left[\left(e^{-4\sqrt{y}}-e^{-4\sqrt{y}\,e^{-s/2}}\right)+4\,\sqrt{y}\,e^{-s/2}e^{-4\sqrt{y}\,e^{-s/2}}\right]\,.
\end{align}

Taking $t=e^{-s/2}$, the integral can be computed explicitly in terms of elementary functions and incomplete $\Gamma$ functions. Wee thus obtain
\begin{align}
    &\mathcal{K}_1(p)=2\int_0^{4\xi^2}dy\,y^3
\int^1_0
dt\,
t^{6+2p}
\Big(
e^{-4\sqrt{y}}
-
e^{-4\sqrt{y}t}
+
4\sqrt{y}\,t\,e^{-4\sqrt{y}t}
\Big)\nonumber\\
&=2\Bigg(\frac{
e^{-8\xi}
\left[
-315
-8\xi
\left(
315
+4\left(
315\xi
+840\xi^2
+1680\xi^3
+2688\xi^4
+3584\xi^5
+4096\xi^6
\right)
\right)
\right]
}{2048(7+2p)}
\nonumber\\[4pt]
&\quad
+\frac{1}{4096}
\Bigg[
\frac{630}{1-2p}
+
\frac{630}{7+2p}
+
\frac{2^{-6p}\xi^{1-2p}\Gamma(7+2p)}{-1+2p}
-315\times 4^{2-3p}\xi^{1-2p}\Gamma(2p,8\xi)
\nonumber\\
&\qquad
-315\times 4^{2-3p}\xi^{1-2p}\Gamma(-1+2p,8\xi)
-315\times 8^{1-2p}\xi^{1-2p}\Gamma(1+2p,8\xi)
\nonumber\\
&\qquad
-105\times 8^{1-2p}\xi^{1-2p}\Gamma(2+2p,8\xi)
-105\times 2^{1-6p}\xi^{1-2p}\Gamma(3+2p,8\xi)
\nonumber\\
&\qquad
-21\times 2^{1-6p}\xi^{1-2p}\Gamma(4+2p,8\xi)
-7\times 2^{-6p}\xi^{1-2p}\Gamma(5+2p,8\xi)-2^{-6p}\xi^{1-2p}\Gamma(6+2p,8\xi)
\Bigg]
\nonumber\\[4pt]
&\quad
+\frac{1}{256}
\Bigg[
\frac{315}{-1+2p}
+\frac{4^{-2-3p}\xi^{1-2p}\Gamma(8+2p)}{1-2p}
\nonumber\\
&\qquad
+315\times 8^{1-2p}\xi^{1-2p}\Gamma(2p,8\xi)
+315\times 8^{1-2p}\xi^{1-2p}\Gamma(-1+2p,8\xi)
\nonumber\\
&\qquad
+315\times 4^{1-3p}\xi^{1-2p}\Gamma(1+2p,8\xi)
+105\times 4^{1-3p}\xi^{1-2p}\Gamma(2+2p,8\xi)
\nonumber\\
&\qquad
+105\times 2^{-6p}\xi^{1-2p}\Gamma(3+2p,8\xi)
+21\times 2^{-6p}\xi^{1-2p}\Gamma(4+2p,8\xi)
\nonumber\\
&\qquad
+7\times 2^{-1-6p}\xi^{1-2p}\Gamma(5+2p,8\xi)
+2^{-1-6p}\xi^{1-2p}\Gamma(6+2p,8\xi)
\nonumber\\
&\qquad
+4^{-2-3p}\xi^{1-2p}\Gamma(7+2p,8\xi)
\Bigg]\Bigg).
\end{align}

Finally, for $\xi\gg 1$, the incomplete $\Gamma$ functions go exponentially fast to zero and the above expression simplifies to
\begin{align}
    \mathcal{K}_1(p)&\simeq 2\frac{p+3}{(2p-1)(2p+7)}\Bigg(\frac{315}{128}-\frac{1}{2^{11+6p}}\frac{1}{\xi^{2p-1}}\Gamma(8+2p)\Bigg)\,,
\end{align}
which is the expression reported in the main body of the paper.

%%%%%%%%%%%%%%%%%%%%%%%%%%%%%%%%%%%%%%%%%%%%%%%%%%%%%%%%%%%%%%%%%%%%%%%%%%%%%%%%%%%
\section{Suppressing the UV cutoff in the momentum integral of eq.~(\ref{eqn:u})}%%
\label{app:cutoff}%%%%%%%%%%%%%%%%%%%%%%%%%%%%%%%%%%%%%%%%%%%%%%%%%%%%%%%%%%%%%%%%%
%%%%%%%%%%%%%%%%%%%%%%%%%%%%%%%%%%%%%%%%%%%%%%%%%%%%%%%%%%%%%%%%%%%%%%%%%%%%%%%%%%%

As discussed in Section~\ref{sec:lap}, all the integrals in  eq.~(\ref{eqn:u}) are convergent even if we send to infinity the UV cutoff appearing in the integral in $dq$ in that equation. It is important to remember that this convergence is an artifact of the UV behavior of the approximate mode function~(\ref{eq:Aapprox}), which is exponentially suppressed and does not provide a good approximation of the exact mode functions (which of course converge to the Bunch-Davies vacuum form, $\sim e^{ike^{-Ht}/H}/\sqrt{2k}$). Nevertheless, since the approximate mode functions naturally cut off precisely to the modes that should be in their vacuum, and therefore irrelevant in any case, the approximation~(\ref{eq:Aapprox}) is commonly used all the way to $k\to\infty$. One could therefore proceed this way also when computing the integrals in eq.~(\ref{eqn:u}).

We show in this Appendix, that, however, the results obtained in this case are qualitatively different from those obtained by keeping the cutoff at $q=4\xi^2$ in the integral in $dq$ in eq.~(\ref{eqn:u}). 

By taking the Laplace transform of eq.~(\ref{eqn:u}) we find that the kernels (which in this case we denote with tildas) take simpler form
\begin{align}
     \mathcal{\tilde{K}}_1(p)&=\frac{315}{64}\frac{3+p}{(2p-1)(2p+7)},\quad
 \mathcal{\tilde{K}}_2(p)=\frac{315}{256}\frac{3+p}{p\,(2p+7)}\,,
\end{align}
and we can write a simple algebraic equation for $U(p)$ as
\begin{align}
U(p)&=\frac{\mathcal{\tilde{N}}(p)}{\mathcal{\tilde{D}}(p)},
\end{align}
where
\begin{align}
    \mathcal{\tilde{N}}(p)&=-\frac{3}{2}\,\kappa\,\Big((p-6)(2p+7)(2p-1)+(p+3)(2p+3)\Big),\nonumber\\
    \mathcal{\tilde{D}}(p)&=p^3(p+3)\,\mathcal{\tilde{F}}(p),\quad \mathcal{\tilde{F}}(p)=(2p-1)(2p+7)+12\,\kappa.
\end{align}

We now see that, while the multiple pole at $p=0$ and the decaying mode at $p=-3$ are present also when the UV cutoff is removed, the infinite zeroes of the function ${\cal F}$ in eq.~(\ref{eq:nd}) are now replaced by only {\it two} zeroes of $\tilde{\cal F}$ at
\begin{align}
    \alpha_\pm=-\frac32\pm\sqrt{4-3\,\kappa}\,.
\end{align}
Moreover, these roots are not even qualitatively similar to those found for finite values of the UV cutoff: for $\kappa>4/3$ (i.e., in the strong backreaction regime) they yield an oscillating  solution, but with decaying amplitude, i.e., a stable mode, whereas for small $\kappa$ they lead to an instability. We have also verified numerically that using the formulae in this appendix we obtain a fit to the actual numerical results (see Figures~\ref{fig:xi25},~\ref{fig:xi30}, and~\ref{fig:xi35}) that is much poorer than that obtained by cutting off the integral at $q=4\,\xi^2$.

%%%%%%%%%%%%%%%%%%%%%%%%%%%%%%%%%%%%%%%%%%%%%%%%%%
%%%%%%%%%%%%%%%%%%%%%%%%%%%%%%%%%%%%%%%%%%%%%%%%%%
%%%%%%%%%%%%%%%%%%%%%%%%%%%%%%%%%%%%%%%%%%%%%%%%%%
%%%%%%%%%%%%%%%%%%%%%%%%%%%%%%%%%%%%%%%%%%%%%%%%%%

\end{document}